\documentclass[sigconf,natbib=false]{acmart}
\usepackage{graphicx} 
\usepackage{subcaption}
\usepackage{xcolor}
\usepackage{amsmath}
\usepackage{xspace}
\usepackage{pifont}
\usepackage{booktabs}
\usepackage{tabularx}
\usepackage{array}
\usepackage{placeins}
\usepackage{afterpage}
\usepackage{float}
\restylefloat{table}
\usepackage[most]{tcolorbox}

\usepackage{enumitem}
\setlist[itemize]{noitemsep, topsep=0pt}
\newtcolorbox{myquote}{
    grow to right by=-4mm,
    grow to left by=-4mm, 
    colback=blue!20!white,
    colframe=blue!80!black,
    overlay={\pgfkeysalso{shadecol=blue!20!white,shadeangle=45}\shade[shading=color wheel] (frame.south west)--(frame.south east)--(frame.north east)--(frame.north west)--cycle;},
    highlight math style= {enhanced, colframe=blue!80!black,colback=blue!20!white,boxsep=0pt},
    boxrule=1pt,
    arc=2mm,
    boxsep=2mm,
    left=1mm,
    right=1mm,
    top=1mm,
    bottom=1mm,
    drop shadow
}

\newtcolorbox{mybox}[3][]
{
  colframe = #2!25,
  colback  = #2!10,
  coltitle = #2!20!black,  
  title    = {#3},
  #1,
}

\newtcolorbox{myboxwithouttitle}[2][]
{
  colframe = #2!25,
  colback  = #2!10,
  coltitle = #2!20!black,  
  #1,
}

\AtBeginDocument{%
  }

\setcopyright{acmcopyright}
\copyrightyear{2024}
\acmYear{2024}
\acmDOI{XXXXXXX.XXXXXXX}
\acmConference[ASIA CCS '24]{Proceedings of the 2024 ACM Asia Conference on Computer and Communications Security}{July 1--July 5, 2024}{Singapore}
\acmBooktitle{Proceedings of the 2024 ACM Asia Conference on Computer and Communications Security (ASIA CCS '24), July 1--July 5, 2024, Singapore}
\acmISBN{978-1-4503-9140-5/22/05}

\newcommand{\M}[1]{\footnotesize\texttt{#1}\normalsize\xspace}

\newcommand{\nProperties}[1]{68}
\newcommand{\nPropertyViolations}[1]{21}
\newcommand{\nVulnerabilities}[1]{3}
\newcommand{\nAttacks}[1]{2}
\newcommand{\nDevices}[1]{17}
\newcommand{\nVulDevices}[1]{14}
\newcommand{\nKnownAttacks}[1]{3}

\RequirePackage[
  datamodel=acmdatamodel,
  style=acmnumeric,
  ]{biblatex}

\begin{document}

\vspace{-7mm}

\title{Segment-Based Formal Verification of WiFi Fragmentation and Power Save Mode}

\author{Zilin Shen}
\email{shen624@purdue.edu}
\affiliation{%
  \institution{Purdue University}
  \city{West Lafayette}
  \state{IN}
  \country{USA}
  }
\author{Imtiaz Karim}
\email{karim7@purdue.edu}
\affiliation{%
  \institution{Purdue University}
  \city{West Lafayette}
  \state{IN}
  \country{USA}
  }
\author{Elisa Bertino}
\email{bertino@purdue.edu}
\affiliation{%
  \institution{Purdue University}
  \city{West Lafayette}
  \state{IN}
  \country{USA}
  }

\begin{abstract}
\looseness = -1
The IEEE 802.11 family of standards, better known as WiFi, is a widely used protocol utilized by billions of users. Previous works on WiFi formal verification have mostly focused on the four-way handshake and other security aspects. However, recent works have uncovered severe vulnerabilities in functional aspects of WiFi, which can cause information leakage for billions of devices. No formal analysis method exists able to reason on the functional aspects of the WiFi protocol. In this paper, we take the first steps in addressing this gap and present an extensive formal analysis of the functional aspects of the WiFi protocol, more specifically, the fragmentation and the power-save-mode process. To achieve this, we design a novel \emph{segment-based} formal verification process and introduce a practical threat model (\emph{i.e.,} MAC spoofing) in Tamarin to reason about the various capabilities of the attacker. To this end, we verify \nProperties{} properties extracted from WiFi protocol specification, find \nVulnerabilities{} vulnerabilities from the verification, verify \nKnownAttacks{} known attacks, and discover \nAttacks{} new issues. These vulnerabilities and issues affect \nVulDevices{} commercial devices out of \nDevices{} tested cases, showing the prevalence and impact of the issues. Apart from this, we show that the proposed countermeasures indeed are sufficient to address the issues. We hope our results and analysis will help vendors adopt the countermeasures and motivate further research into the verification of the functional aspects of the WiFi protocol.

\end{abstract}



\maketitle
\newcommand{\elisa}[1]{\textbf{\color{red}Elisa: #1}}
\section{Introduction}
WiFi is a ubiquitous wireless networking technology that allows devices to connect to the Internet and connect to each other~\cite{nazir2021survey}. It operates based on the 802.11 family of standards, which indicate the specifications~\cite{specification} and protocol for wireless local area networks (WLANs). Currently, WiFi protocol has several versions, 802.11n/ac/ax. 

\looseness = -1
Because of its wide use in all application domains, the security of WiFi communication is critical. One important approach to address such a requirement is to formally verify the WiFi protocol to
detect vulnerabilities in its design. Formal verification approaches have been proposed as part of past work~\cite{cremers2020formal,basin2018formal,shiformal,hussain2018lteinspector,hussain20195greasoner,wu2022formal,jacomme2023comprehensive,cremers2019component,zhou2018static}, focusing on different wireless communication protocols, such as LTE~\cite{ghosh2010lte}, 5G~\cite{gupta20155gsurvey}, and Bluetooth~\cite{haartsen2000bluetooth}.
Cremers et al.~\cite{cremers2020formal} verified the WiFi protocol with a focus on WPA2 handshakes, such as four-way handshakes and group-key handshakes. They verified the KRACK attack~\cite{vanhoef2017key} using Tamarin~\cite{meier2013tamarin}, a \emph{cryptographic verifier}. 
However, their verification scope is only the WiFi handshake, which limited their detected attacks to KRACK. 
Recently, vulnerabilities have been identified ~\cite{vanhoef2021fragment, schepers2023framing} 
in the fragmentation and power save mode (PSM) components of the 
WiFi protocol~\cite{vanhoef2021fragment, schepers2023framing}. Such vulnerabilities may lead to the leakage of user-sensitive information. 

Because the WiFi protocol is so widely used, attacks on the WiFi functional components affect billions of devices. However, to date, no formal verification approach has been proposed to analyze the security of the WiFi functional components.
Therefore, in this paper, we address the problem of designing a formal approach to verify the functional components of the WiFi protocol--concretely, the fragmentation and PSM functional components of the protocol.
Fragmentation and PSM are both closely related to the frame buffer mechanism of the WiFi protocol. The reason is that the buffered units are stored in plain text in the buffer, which may cause severe vulnerabilities. 
Thus, we focus on the formal verification of these two parts in our work as the first steps toward the verification of the functional components of the WiFi protocol.


The design of a formal approach for the verification of 
WiFi functionalities require addressing the following
challenges:
{\itshape (C1): Modelling multi-protocol interactions:} The WiFi protocol specification~\cite{specification} has more than 4,000 pages and has complex contents. Modeling the functional protocols of WiFi also requires considering the WiFi security protocols, such as RSNA (Robust Security Network Association).
Therefore, the approach must account for the interaction between the functional components (such as fragmentation and PSM) and the security protocols. 
{\itshape (C2): Modelling the MAC spoofing threat model:} There is a wide range of attacks~\cite{vanhoef2017key,schepers2023framing} against WiFi that leverage MAC spoofing.
The current formal verification approaches cannot reason about this important WiFi threat vector. 

In this paper, we build a comprehensive Tamarin~\cite{meier2013tamarin} model for WiFi protocol, including both \textit{fragmentation} and \textit{PSM}. For the fragmentation, we model the entire process from the sender transmitting a 
MSDU (MAC Service Data Unit) to the receiver accepting it, including the case when the MSDU is long and thus must be fragmented during transmission. In our model for the fragmentation process, we also include the security encapsulation method of the WiFi channel.
For the PSM, we build a model that represents the cycle alternating process, which includes the conversion of the station's active state and power save state. We also include in the model the message transmission and buffering 
during this conversion process.
To address the complex interaction between the WiFi functional protocols and RSNA security protocols, we design a novel approach, referred to as the \emph{segment-based} Tamarin model, that extends Tamarin based on the notion of segmentation. 
Compared to the previous work that divide the Bluetooth protocol into different linear procedures and model them as modules~\cite{wu2022formal}, in our case, the functional protocol and security protocol interaction create complex protocol interactions. 
To this end, we carefully divide the whole model into several segments, where each protocol is a segment, and then design how these segments interact with each other. 
When we need a particular segment in the model, we directly call the functions of this segment. Our idea is motivated by object-oriented programming and the concept of \emph{encapsulation}. We create several segments as classes and when needed a user can call these classes to instantiate the object as a formal model. Following the principle of encapsulation, the internal mechanisms of the segment are kept private and invocation mechanisms are made public to achieve interaction with the other protocols.
To address the MAC spoofing threat model challenge, we add a new attack entity in our Tamarin code to model an attacker with MAC spoofing capabilities.

We design and implement our model with Tamarin, which includes the fragmentation and PSM components. We verify \nProperties{} properties from the WiFi protocol specification with our model. 
We identify \nPropertyViolations{} properties violation from the verification and then 
conclude that  \nVulnerabilities{} of those are vulnerabilities.
From the identified vulnerabilities, we propose \nAttacks{} attacks. We then evaluate the attacks on \nDevices{} commercial devices spanning various vendors and WiFi generations.
Notably, \nVulDevices{} devices are found vulnerable, which proves that most devices are affected by the issues we have identified. We further discuss countermeasures to mitigate these issues.
Beyond our findings, \nKnownAttacks{} existing attacks~\cite{vanhoef2021fragment,schepers2021framework} are detected by our model. Additionally, we have developed a patched version model that proves resistant against these known attacks.

Our paper's contributions are as follows:
\begin{itemize}

    \item To the best of our knowledge, we are the first to formally verify the functional components of the WiFi protocol. More specifically, we verify the fragmentation and PSM of the protocol, which have been shown to be vulnerable in previous works; attacks on these functions can have severe security and privacy impacts. 
    \item We design and implement a segment-based analysis (inspired by object-oriented programming and the principle of encapsulation) to create the formal model in Tamarin. Furthermore, we include a new threat model, which expands the verification scope and allows us to reason about the various attacker capabilities and a wide range of properties.
    \item We verify \nProperties{} extracted properties from WiFi protocol specification. We find \nVulnerabilities{} vulnerabilities from the verification and propose \nAttacks{} new issues. These vulnerabilities and attacks affect \nVulDevices{} commercial devices out of \nDevices{} cases, showing the prevalence and impact of the issues. Our model also detects \nKnownAttacks{} existing attacks. We then implement and verify the patched-version Tamarin model, which proved resistant to existing and new attacks.
\end{itemize}

\noindent \textbf{Open-source.} Artifacts related to this project's formal verification and analysis are open-sourced on Github~\cite{github}.

\section{Background}
This section provides an overview of the fragmentation, defragmentation, and PSM of the WiFi protocol and a short introduction to the Tamarin prover.

\subsection{Fragmentation and Defragmentation}

\begin{figure}[h]
  \centering
  \includegraphics[width=\linewidth]{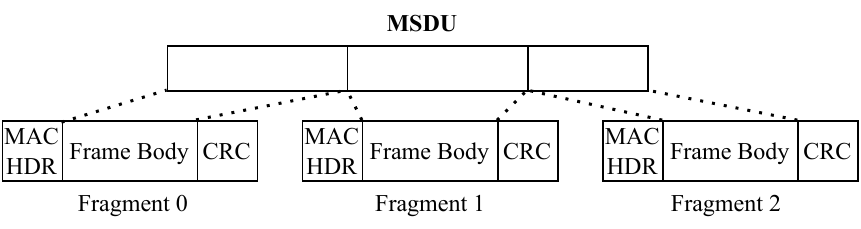}
  \caption{Outline of the fragmentation process based on the WiFi protocol specification~\cite{specification}. }
  \label{fig:fragmentation-background}
\end{figure}

Fragmentation, a mechanism introduced with the early 802.11 wireless protocol, was designed to decrease the probability of interference during long-frame transmission. To accomplish this, fragmentation breaks a long MSDU into smaller fragments for sequential transmission, as illustrated in Fig~\ref{fig:fragmentation-background}. Then MAC header and CRC (Cyclic Redundancy Check) code are added to the fragment, transforming the fragment into an MPDU (MAC Protocol Data Unit). MPDU is the transmission unit in the WiFi channel. Conversely, defragmentation is a mechanism that reassembles those fragments. 

\begin{figure}[h]
  \centering
  \includegraphics[width=\linewidth]{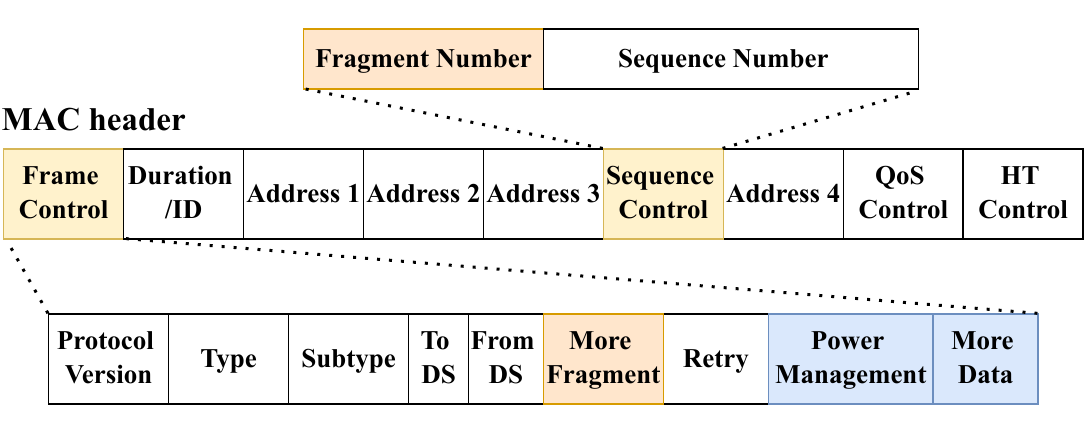}
  \caption{The MAC header and details about Frame Control and Sequence Control. The orange area indicates the header fields related to fragmentation. The blue area indicates the header fields related to PSM.}
  \label{fig:mac-header}
  
\end{figure}

In the fragmentation mechanism, there are two 
critical flags in the packet MAC header (see the 802.11 specifications \cite{specification}). 
These flags are the \textit{More Fragment} flag in the Frame Control field and the \textit{Fragment Number} field in the Sequence Control area (see Fig~\ref{fig:mac-header}).
When the \textit{More Fragment} flag is set to 0, 
it indicates that the current frame is either the last fragment of a set or it is a standalone frame that has not been fragmented. Conversely, when a \textit{More Fragment} flag is set to 1, it indicates that the current frame is part of a larger set, denoting that more fragments follow.
The \textit{Fragment Number} field is equally significant in the fragmentation mechanism. It indicates the serial number of each fragment in a series. When a frame is divided into multiple fragments, 
the first fragment is assigned a \textit{Fragment Number} equal to 0, the second fragment a \textit{Fragment Number} equal to 1, and so forth for all the subsequent fragments. Basically, the $n^{th}$ fragment is assigned a \textit{Fragment Number} equal to n-1.
Because the \textit{Fragment Number} comprises 4 bits, the fragment number ranges from 0 to 15.

In the fragmentation process, one MSDU can be divided into multiple MPDUs. The receiver's task is then to reassemble these MPDUs back into the original MSDU. 
The Sequence Control and Frame Control fields are utilized by the receiving station (STA) to accurately conduct this reassembly. The \textit{More Fragment} flag within the Frame Control field serves as an essential indicator during this process. When an MPDU with a \textit{More Fragment} flag of 0 is received, it indicates the end of a fragmented series; the receiver then proceeds to reassemble all the MPDUs with the same \textit{sequence number}. The order of reassembly adheres to the increasing \textit{fragment number} order, ensuring that the original MSDUs' integrity and structure are maintained.

\subsection{Power Save Mode}
Power Save Mode (PSM) is a fundamental energy-saving feature of the 802.11 protocol~\cite{rozner2010powersave}. For a normal STA, 
there are two operational modes, namely active mode and power save (PS) mode. The key idea behind the energy-saving feature is managing the operational state of the STA to optimize power usage. When the STA
enters the PS mode, the corresponding downlink data is held or buffered at the Access Point (AP), instead of being transmitted immediately. The 802.11 protocol defines the notion of a Bufferable Unit (BU), as a data unit that can be buffered with a PS mechanism. When the STA wakes up, it sends a request to ask the AP for the stored data. In response to the request, the AP forwards the buffered data units to the STA. 

The \textit{Power Management} subfield and \textit{More Data} subfield in the Frame Control Field of the MAC header are related to PSM, as detailed in Section 9.2.4 of the WiFi protocol official documentation~\cite{specification}; they are shown as the blue area in Fig~\ref{fig:mac-header}. 
The \textit{Power Management} subfield serves as an indicator of an STA's power management mode. A value of 1 in this subfield indicates that the STA will transition into the PS state. 
Conversely, a value of 0 indicates that the STA will maintain an active state.
The \textit{More Data} subfield indicates that additional BUs are stored for that STA at the AP. If the \textit{More Data} subfield is set to 1, it indicates that the AP holds at least one additional BU for that STA. 


\begin{figure}[htb]
  \centering
  \includegraphics[width=0.8\linewidth]{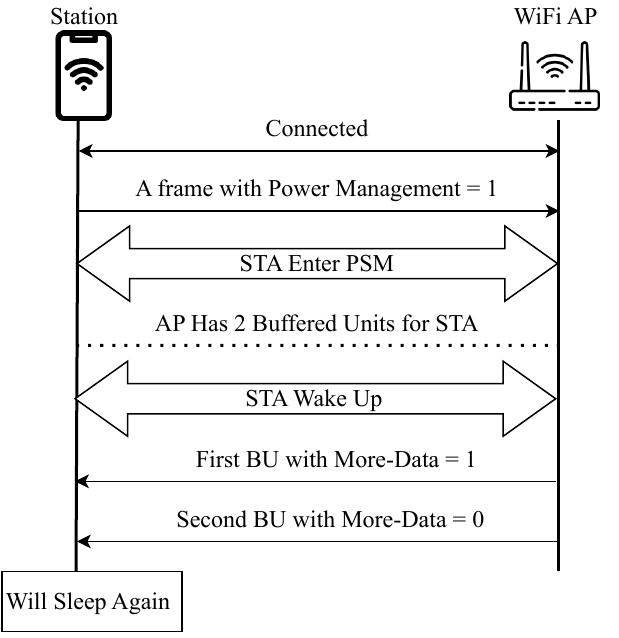}
  \caption{The PSM process. It includes the STA entering power save mode and then waking up.}
  \label{fig:PSM-process}
  
\end{figure}

The entire process of an STA transitioning into PS mode and subsequently waking up is illustrated in 
Fig~\ref{fig:PSM-process}. The figure shows how the STA transitions to the power save mode and returns to the active mode. This process has the following steps:

\begin{itemize}
\item When the STA wants to change from the active mode to the PS mode, it informs the AP by completing a successful frame exchange initiated by the STA. The \textit{Power Management} subfield of the frame from STA should be 1 to declare it will enter PS mode. 
\item Upon receiving an ACK message from the AP, the STA enters the PS mode. During this period, the AP is required to buffer any incoming BUs intended for the STA.
\item After the STA wakes up, the STA will send a PS-poll message to the AP asking for a buffered message.  As a response, the AP sends the first buffered message to the STA with the \textit{More Data} field set to 1, indicating that more buffered data remains to be sent.
\item Once the STA receives this initial buffered message, it checks the value of the \textit{More Data} field. If the value is equal to 1, it sends another PS-poll message to request the next buffered message. As this is the final buffered message, the AP transmits it with the \textit{More Data} subfield set to 0, indicating the end of the buffered data series.
\end{itemize}

\subsection{The Tamarin Prover}
The Tamarin prover is an advanced tool to facilitate automated, symbolic analysis of security 
protocols~\cite{meier2013tamarin}. Tamarin generalizes the backward search to enable protocol specification by multiset rewriting rules~\cite{durgin2004multiset}, property specification in a guarded first-logic logic, and reasoning modulo equational theories. As a result, Tamarin can handle protocols with complex control flow and loops, complex security properties, and equational theories. Given these capabilities, we
select Tamarin as our tool of choice for modeling the fragmentation and the PSM components of the WiFi protocol. 

To formalize the security protocol with Tamarin, we represent the protocol as a collection of multiset rewriting rules. Such as:

\begin{myboxwithouttitle}{blue}
		\small
    \textsf{[ SenderState (senderID, seqNum, nonce, key),} \newline
    \textsf{SenderMessage (message)]} \newline 
  --\textsf{[ SenderSendFragment (senderID, message, seqNum, nonce)]} \newline 
  ->\textsf{[ Out(<seqNum, nonce, senc(message,key)>)]}
\end{myboxwithouttitle}

This rule specifies that when a sender identified by an ID and sequence number has a message to transmit, it encrypts the message using a symmetric encryption function and sends it out.
The terms \textsf{SenderState(senderID,seqNum,nonce,key)}, and \textsf{SenderMessage(message)} are referred to as \textit{left-hand facts}, representing the initial state before the rule is applied.
\textsf{Out(<seqNum, nonce, encryptedMsg>)} is referred to as the \textit{right-hand fact}, symbolizing the resultant state after the rule's application. The term \textsf{SenderSendFragment(senderID,message,seqNum, nonce)} is referred to as the \textit{action fact}, indicating the action to be executed.
The term \textsf{senc} is a built-in function in Tamarin for a symmetric encryption function~\cite{bellare1997concrete}.  Tamarin provides support for defining custom functions, allowing one to specify the semantics of these functions using equations. We leverage this feature to model the fragmentation and defragmentation processes, enabling a precise and faithful representation of these complex aspects of the WiFi protocol.

Once the entire protocol is specified using multiset rules,
we can use first-order logic formulas~\cite{barwise1977first-order} to represent security properties. These properties may encompass confidentiality, authentication, integrity, and more, serving as metrics for verifying the security of the protocol. For instance, consider the following first-order logic formula:
\begin{myboxwithouttitle}{blue}
    \small
    \textsf{All \#j msg. ReceiverRecMsg(msg) @j} \newline
    ==> \newline
    \textsf{(Ex \#i. (SenderSendMsg(msg) @i) \& i<j)} 
\end{myboxwithouttitle}
The formula specifies that, for all protocol traces, if the receiver obtains a message at time \textit{j}, there should exist a time \textit{i} at which the sender sends the message, with time \textit{i} occurring before \textit{j}. This property helps in verifying the integrity of the message, ensuring the data has not been modified during transmission. 

One can also incorporate restrictions in a Tamarin model, which serve to define specific constraints for the model. For instance:
\begin{myboxwithouttitle}{blue}
    \small
    \textsf{restriction Equality:} \newline 
      \textsf{"All x y \#i. Eq(x,y) @ i ==> x = y"}
\end{myboxwithouttitle}
This restriction specifies the condition of equality, allowing one to use \textsf{Eq(x,y)} in the action fact to represent an equation constraint between two items. This restriction is useful for integrity verification, that is, to verify that the transmitted MIC (Message Integrity Code) is the same as the one calculated MIC on the receiver side. 

\section{Overview}

\label{sec:overview}
In this section, we first discuss the challenges of our analysis methodology and our approaches to address these challenges.
We then outline our methodology's workflow, followed by a high-level description of each workflow step.

\subsection{Challenges}
In what follows, we discuss the challenges related to the formal verification of WiFi fragmentation and PSM and the approaches to addressing them.

\begin{figure}[h]
  \centering
  \includegraphics[width=0.7\linewidth]{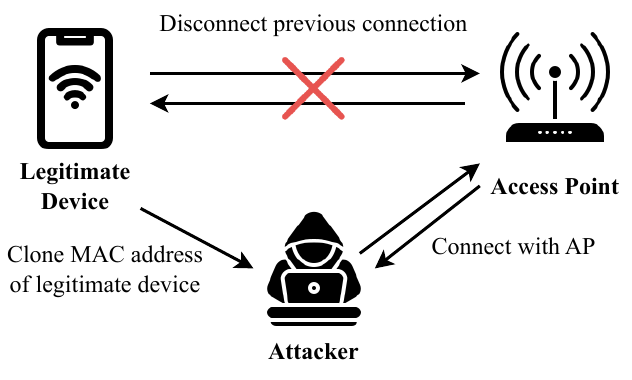}
  \caption{Overview of MAC spoofing attack. }
  \label{fig:mac-spoofing}
  
\end{figure}

\noindent \textbf{(C1) Modeling Multi-Protocol Interactions:} The Finite State Machine (FSM) construction and property extraction require a thorough analysis of the complex WiFi protocol specification~\cite{specification}, which extends over 4,000 pages. 
As the fragmentation and PSM components interact with the security protocol, in our model, we need to consider the interplay between the RSNA protocols (such as CCMP and GCMP) and the fragmentation protocol. For instance, the authentication of a message that occurs either prior to or following the fragmentation of the frame will result in a variety of security outcomes. For our patched version model, we need the authentication process both before and after the fragmentation. This means the authentication of the frame and a series of fragments are all needed. This patched model makes the interaction complex and bi-directional.
As for the PSM part, the model must represent the interaction between the SA (Security Association) and the PSM functions. When a STA enters the PS mode, the AP buffers the packets destined for the device. Hence, the PSM model must account for how buffered units are protected and transmitted under the considered security protocol.


To address this challenge, we first carefully analyze these interactions and use a segment-based model.
We partition the whole model into several segments and represent how each segment interacts with each other. For building the fragmentation model, we need to consider both the fragmentation functional segment, the security encapsulation (CCMP .etc) segment, and their interactions. 
In our implementation, we first design and implement the security encapsulation segment. When the fragmentation model needs the messages to be security encapsulated, we can then directly invoke the functions in the security encapsulation segment.
To achieve these, we use the m4 language\cite{kernighan1977m4}, which acts as an intermediate generator of Tamarin code.  Detailed information is provided in Section~\ref{sec:model-construction}.


\noindent \textbf{(C2) Modeling the MAC Spoofing Threat Model:} 
A MAC spoofing attack~\cite{guo2006macspoofing} allows a malicious entity to impersonate a legitimate node by using a falsified MAC address. The AP cannot concurrently maintain a connection with more than one device with the same MAC address. Therefore, when MAC spoofing happens, the connection between the AP and the legitimate device disconnects first. Then the attacker uses the MAC address of the legitimate device to establish a connection with the AP (see Fig~\ref{fig:mac-spoofing}). Because the management and control frame not being protected by the existing security protocol, the attacker can launch MitM (Man-in-the-Middle)~\cite{conti2016mitm} and DoS (Denial-of-Service)~\cite{peng2007dos} attacks~\cite{liu2019mac}.  
In addition, the AP relies on MAC addresses to encode critical information about each device, such as the security association or the bitmap indicating whether a device is active or in sleep mode. Consequently, a MAC spoofing attack can introduce severe issues, particularly related to fragmentation and PSM. 
Therefore, we need to add the MAC spoofing to the attacker's capabilities of the model so that these severe issues can be detected.
In the meanwhile, incorporating MAC spoofing considerably broadens the scope of our model's verification.

Modeling MAC spoofing in Tamarin presents a challenge. Tamarin's default threat model is the Dolev-Yao model~\cite{dolev1983security}, where the built-in facts \textsf{Out(message)} and \textsf{In(message)} are under the assumption that the communication channel 
operates under the Dolev-Yao model, thereby eliminating the need for user-defined attacker functions.
To simulate MAC spoofing, we introduce a new entity—an attacker—and define the MAC spoofing attacker's capabilities. According to the workflow shown in Fig~\ref{fig:mac-spoofing}, we add a new attacker entity and construct an FSM~\cite{lee1996fsm} representing the attacker. 
We define the capabilities of the MAC spoofing attacker in our model, which includes tampering with the security association and fragment queue of the AP. The details are shown in Section~\ref{sec:threat-model}.

\begin{figure}[htb]
  \centering
  \includegraphics[width=0.85\linewidth]{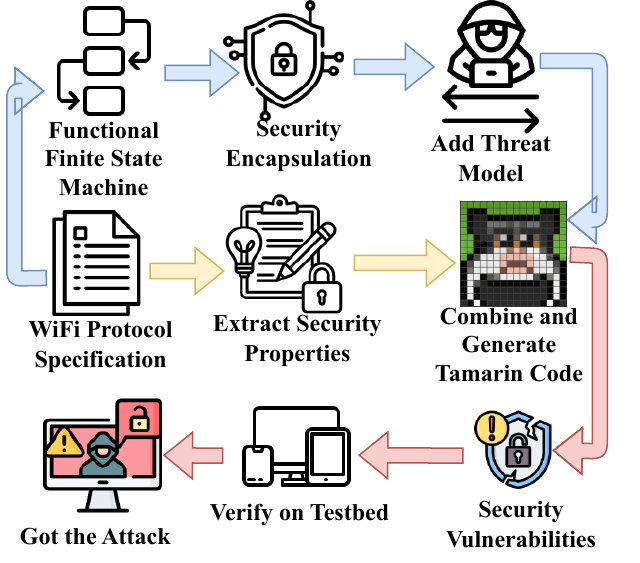}
  \caption{Overview of the workflow. There are three main phases: (1) model construction (blue arrow); (2) properties extraction (yellow arrow), and (3) properties and testbed verification (red arrow). The first phase builds the model from the protocol specification, and the second phase extracts the security properties from the protocol specification. The third phase verifies the extracted properties and tests vulnerabilities on the testbed.}
  \label{fig:workflow}
\end{figure}

\subsection{Workflow}

The workflow of our methodology comprises three core phases: model construction, extraction of security properties, as well as security properties and testbed verification (see Fig~\ref{fig:workflow}).

\noindent \textbf{Model Construction:} It comprises four distinct processes: (i) analysis of the specification to derive the FSM, (ii) addition of the security encapsulation methods, (iii) characterization of the threat model to define attacker capabilities, and (iv) translation of the derived FSM model into Tamarin code.

To construct a model of the fragmentation and PSM components, it is necessary to thoroughly examine the official WiFi protocol specification documentation~\cite{specification}. 
Subsequently, we construct FSMs for each entity of the model, which is necessary for Tamarin model construction. 

Upon completing the FSMs of each entity, we consider the security encapsulation method. The functional FSM part determines when to send a message and what message to send. 
Before the message is sent to the channel, the message should be encrypted, authenticated, and the integrity code should be added to ensure integrity, which process is called security encapsulation.
How the message is securely encapsulated will determine the security of the communication process.
Our model needs a complex interaction of functional segments and security segments. This is the key insight for proposing a segment-based design to construct the model.

Then, we consider the threat model. 
Within the Tamarin context, the threat model outlines the capabilities of potential adversaries in the system. We consider two threat models in our system, the Dolev-Yao threat model~\cite{dolev1983security} and the MAC spoofing threat model. 

Finally, we combine the threat model and FSM from protocol specification together to translate them into Tamarin~\cite{meier2013tamarin} code to facilitate subsequent verification.

\noindent \textbf{Security Properties Extraction: } Security properties are specific assertions or conditions that one aims to verify within the protocol. The properties can include various security aspects, including secrecy, authentication, integrity, and privacy. The extraction process involves two critical steps: first, security properties are extracted from the documentation, and subsequently, these properties are translated into first-order logic within the Tamarin code.  This part is shown as yellow arrows in Fig~\ref{fig:workflow}.

\noindent \textbf{Verifying Properties and Testbed Verification: } This phase involves verifying the extracted properties and the vulnerabilities identified by the Tamarin and testing the issues on the testbed.
Following the model construction and property specification from the specification, Tamarin can be utilized to verify all properties. While some properties will be verified, others will be falsified. These property violations can then be concluded as several vulnerabilities.
These vulnerabilities and new attacks generated are then verified with a WiFi testbed. Furthermore, we test several commercial devices to evaluate the impact.
Upon successful testbed verification, we identify actual attacks. 

\section{Design}
\label{sec:design} 
This section details the processes for building the model and identifying the security properties. Details about properties verification and testbed evaluation are given in Section~\ref{sec:evaluation}.
\begin{figure*}[htbp]
    \centering
    \begin{subfigure}{0.5\textwidth}
        \centering
        \includegraphics[width=0.8\linewidth]{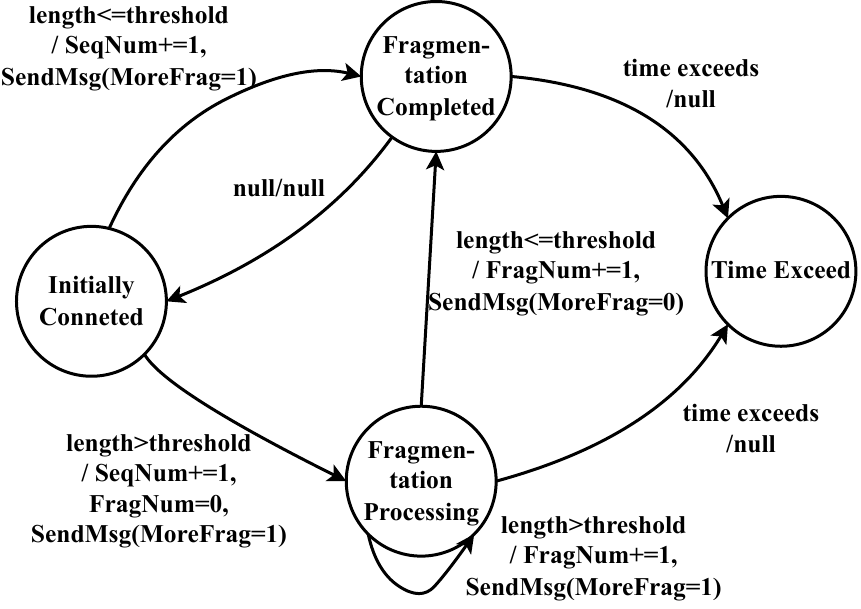}
        \caption{The FSM of fragmentation process}
        \label{fig:frag-FSM}
    \end{subfigure}%
    \begin{subfigure}{0.5\textwidth}
        \centering
        \includegraphics[width=0.8\linewidth]{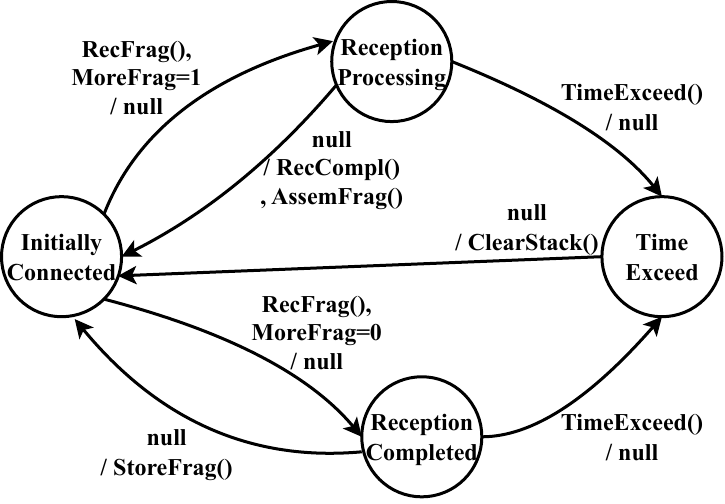}
        \caption{The FSM of defragmentation process}
        \label{fig:defrag-FSM}
    \end{subfigure}
    \caption{The FSMs related to WiFi fragmentation component. The fragmentation FSM is on the sender side. The defragmentation FSM is on the receiver side.}
\end{figure*}
\subsection{Threat Models}
\label{sec:threat-model}
We consider two threat models: the Dolev-Yao model~\cite{dolev1983security} and the MAC spoofing model, derived from the MAC spoofing attack~\cite{liu2019mac}. 

\noindent \textbf{Dolev-Yao Model:} The Dolev-Yao model is the most commonly used threat model 
in security verification. Under this model, adversaries possess the capability to overhear, intercept, and generate any message, with their only constraint being the guarantees of the used cryptographic techniques. The Dolev-Yao model is the default threat model in Tamarin~\cite{meier2013tamarin}.
We can use the default \textsf{Out(message)} and \textsf{In(message)} in Tamarin to represent the sending and receiving of messages via the channel that adheres to the Dolev-Yao threat model.

\noindent \textbf{MAC Spoofing Model: } A MAC Spoofing attack allows the adversary to impersonate a legitimate node by using a falsified MAC address. To simulate MAC spoofing, we introduce a new entity--a MAC spoofing attacker--and define the attacker's capabilities. 
The AP primarily utilizes the MAC address as a unique identifier for each client device. 
Subsequently, under the MAC spoofing threat model, the attackers can alter the information of the impersonated device stored in AP.
The MAC spoofing attacker can thus manipulate the following stored information:
\begin{itemize}
    \item the bitmap that indicates whether each device is in active mode or PSM;
    \item the buffered units that use the MAC address to indicate the source device;
    \item the security association that includes the pairwise key. 
\end{itemize}
Therefore, we utilize Tamarin to implement these capabilities of attacker to include the MAC spoofing attacker into our model.

\subsection{Model Construction}
\label{sec:model-construction}
\begin{figure}[h]
  \centering
  \includegraphics[width=\linewidth]{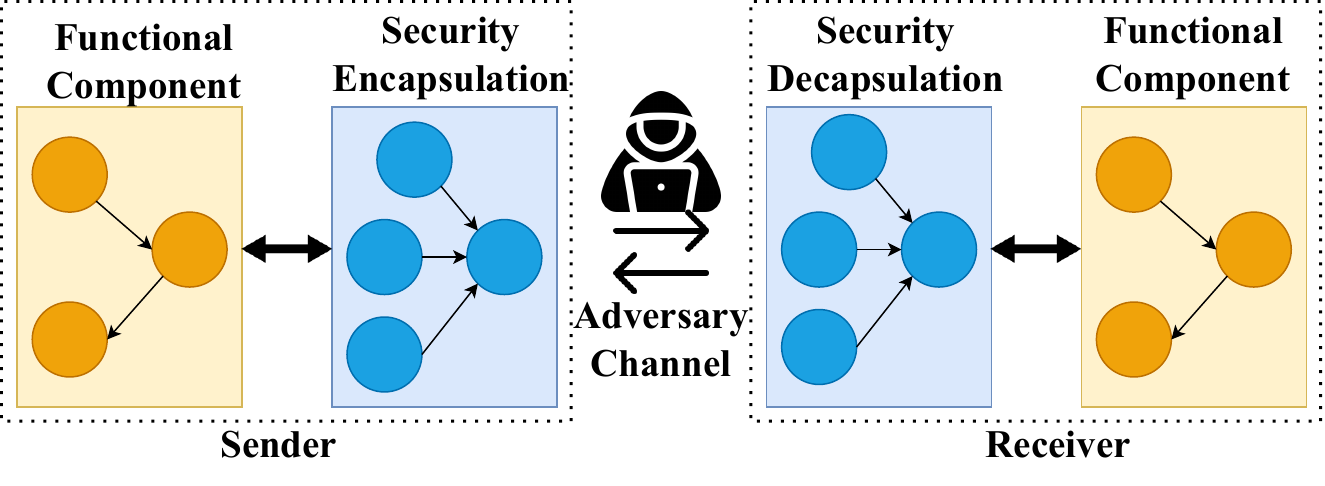}
  \caption{Model construction structure. WiFi functional components (yellow areas) interact with security encapsulation and decapsulation components (blue areas). }
  \label{fig:frag-whole-process}
\end{figure}

The model construction comprises four main steps: analyzing the WiFi protocol specification to derive the FSM, combining the security encapsulation component, adding the threat model, and translating the result into Tamarin.
Note that we use a segment-based methodology to address the interaction across WiFi functional and security components. 
A high-level view of the interactions between the functional components, security methods, and threat model is given in Fig.~\ref{fig:frag-whole-process}. Because we focus on the fragmentation and PSM components of WiFi, the functional components include these two processes. 
Security encapsulation and decapsulation refer to WiFi RSNA. 
In the following, we provide details on fragmentation and defragmentation model construction, PSM model construction, as well as security encapsulation and decapsulation.

\subsubsection{Fragmentation and Defragmentation Model Construction}
To construct the model, the first important step is to examine the WiFi specification ~\cite{specification} to obtain the FSMs~\cite{lee1996fsm} of the fragmentation and defragmentation parts. 
Two entities are involved in fragmentation and defragmentation: the sender, responsible for the fragmentation process, and the receiver, responsible for reassembling the fragments. 
We thus formulate two FSMs for the sender and receiver, respectively. 

The FSM representing the fragmentation process at the sender's end is shown in Fig.~\ref{fig:frag-FSM}. We examine \textbf{Section 10.4 MSDU and MMPDU fragmentation} of the WiFi specification for the FSM design.
The fragmentation FSM has four distinct states: 
\M{Initially Connected}, \M{Fragmentation Processing}, \M{Fragmentation Completed}, and \M{Time Exceed}. The meaning of each state is shown below. 
\begin{itemize}
    \item \M{Initially Connected}: The sender has connected with the receiver and has some MSDUs ready to send.
    \item \M{Fragmentation Processing}: The MSDU length is longer than the threshold; the sender needs to separate the frame into several fragments. 
    \item \M{Fragmentation Completed}: The MSDU length is below the threshold, with no need for fragmentation, or it is  
    the last fragment.
    \item \M{Time Exceed}: Current 
    MSDU's transmission time threshold has been exceeded.
\end{itemize}

Following the specification, the frame length should not be larger than \textit{dot11FragmentationThreshold}, which we refer to as the threshold in this paper.
When the sender is at \M{Initially Connected} state with several MSDUs to transmit, whether the fragmentation is needed depends on the length of the MSDU frame.
If the length is fewer than the threshold, it will directly enter \M{Fragmentation Completed} state and send the message out; if the length is more than the threshold, it will enter \M{Fragmentation Processing} state and separate the frame to several fragments. Then, the fragmentation process will continue until the remaining frame length falls below the threshold.



Then, we move to the rules about \textit{fragment number}. 
After the examination of the specification, in our fragmentation FSM, the \textit{fragment number} should start at 0 and increment by 1 each time. 
Another fragmentation-related variable, the \textit{sequence number}, 
is identical for every frame. As for the \textit{More Fragment} flag, indicating the presence of additional fragments for the current frame, 
all fragments (except the last one) should have their \textit{More Fragment} flag designated as 1 throughout the fragmentation procedure. 

Now, moving on to the FSM of the defragmentation process, shown in Fig.~\ref{fig:defrag-FSM}, the FSM is extracted from~\textbf{Section 10.5 MSDU and MMPDU defragmentation} of WiFi protocol specification ~\cite{specification}. 
There are four states in the defragmentation FSM:
\begin{itemize}
    \item \M{Initially Connected}: The receiver has connected with the sender and listening for messages. 
    \item \M{Reception Processing}: The receiver is receiving several fragments with the More Fragment flag equal to 1.
    \item \M{Reception Completed}: The receiver receives the fragment with the More Fragment flag equal to 0.
    \item \M{Time Exceed}: The timer maintained on the receiver side exceeds.
\end{itemize}

Upon receiving an MPDU or fragment with the \textit{More Fragment flag} set to 1, meaning more fragments will arrive, the receiver transitions to the \M{Reception Processing} state. In this state, the received fragments are buffered for subsequent reassembly.
When a fragment arrives with the \textit{More Fragment flag} equal to 0, it indicates the termination of a fragment series. Consequently, the receiver transitions to the \M{Reception Completed} state. Here, buffered fragments sharing identical sequence numbers and originating from the same source MAC address are systematically reassembled based on their respective fragment numbers.
Like the fragmentation FSM, the defragmentation process has a timing constraint. If a timer for a specific MSDU or frame exceeds the threshold
, the receiver should discard all buffered fragments and any subsequent fragments belonging to the frame.
\vspace{-3mm}
\subsubsection{PSM Model Construction}

We start by presenting the FSMs associated with the PSM. This will be supported with references from the WiFi protocol specification. We then present some simple examples using Tamarin code to illustrate our approach.

\begin{figure}[htb]
    \centering
    \begin{subfigure}{0.5\textwidth}
        \centering
        \includegraphics[width=0.8\linewidth]{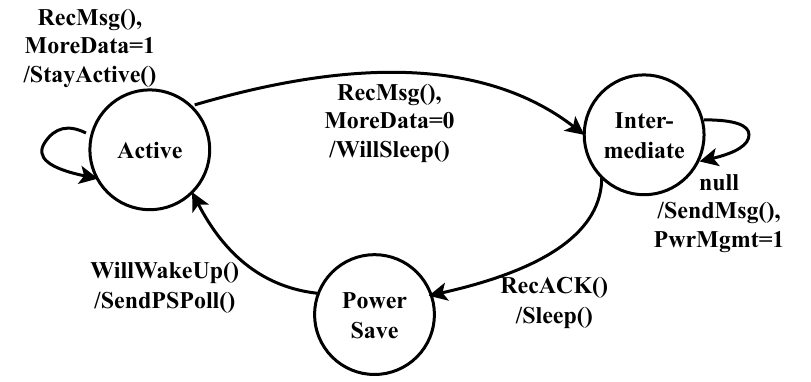}
        \caption{The FSM on Station side}
    \end{subfigure}

    \begin{subfigure}{0.5\textwidth}
        \centering
        \includegraphics[width=0.8\linewidth]{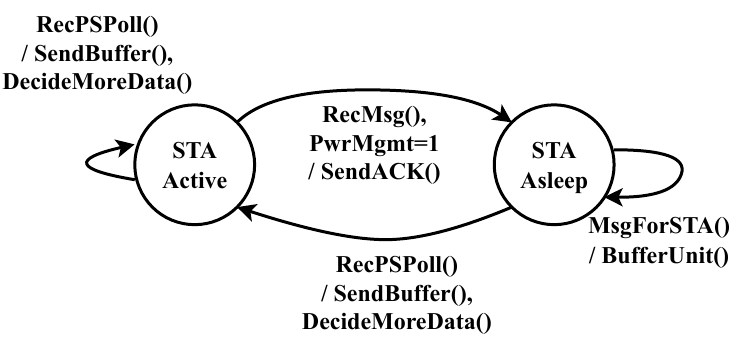}
        \caption{The FSM on Access Point side}
    \end{subfigure}
    \caption{PSM component include the STA FSM and AP FSM.}
    \label{fig:psm-sta}
\end{figure}
We follow \textbf{Section 11.2 Power Management} of the WiFi protocol specification ~\cite{specification} to build the PSM model. In this model, we consider two identities, STA and AP. Therefore, we build two FSMs for them, shown in Fig.~\ref{fig:psm-sta}. 

An STA can be in one of two power states: active and power save. In addition, one intermediate state indicates that the STA sends the entering power save mode message to the AP but hasn't received the ACK from the AP. Therefore, there are three states in the STA FSM: 
\begin{itemize}
    \item \M{Active}: STA is fully powered.
    \item \M{Intermediate}: The STA wants to sleep and thus sends a message to the AP, but has not received the ACK.
    \item \M{Power Save}: STA is not able to transmit or receive and consumes very low power. 
\end{itemize}
According to the specification, when the STA wants to enter a power save state, it should first send a message with \textit{Power Management (PwrMgmt)} flag equals to 1. The state of the STA FSM will change from \M{Active} to \M{Intermediate}, and the STA will wait for the ACK message from the AP. 
When the STA receives the ACK message from the AP, it will enter the power save state; in the FSM, the state will change from \M{Intermediate} to \M{Power Save}. 

When the STA wakes up from power save mode, it signals the AP about this state change by sending a PS-poll message. 
This message also requests buffered messages stored in AP while the STA is in power save mode. 
In the context of our FSM, this transition is represented as a change from the \M{Power Save} state to the \M{Active} state.
Upon receiving the buffered frame from the AP, the STA checks the \textit{More Data} flag. The value of this flag indicates the AP's storage status.
If the flag is set to 1, it indicates that additional buffered messages are stored in the AP. 
Conversely, if the flag is set to 0, it indicates that no more buffered messages remain in the AP. Recognizing this, the STA will transition back to its power save state, starting a new power-saving cycle.

Then we move to AP PSM; there are two states in AP PSM:
\begin{itemize}
    \item \M{STA Active}: the connected STA is active.
    \item \M{STA Asleep}: the connected STA is in power save mode.
\end{itemize}
When the AP receives the message with \textit{Power Management frag} equals to 1, the AP will change the bitmap that saves the power state of STA from active to power save and send the ACK message back to STA. In the AP FSM, the state will change from \M{STA Active} to \M{STA Asleep} state. 

When the AP is aware that some STA is in power save mode, the AP will buffer the messages destined for this STA. Meanwhile, the information on MAC service, such as the source MAC address and destination MAC address should be maintained. 
When the AP receives the PS-poll message from the STA asking for buffered messages, the AP will send the STA 
the buffered message encrypted by current security association information (which includes the pairwise key). If the message is not the last in the buffer, the \textit{More Data} flag should be 1 to indicate there are more messages in the buffer.

We provide a Tamarin code example showing how to implement the "AP knows that the STA will enter power save state and sends the ACK back" in Appendix~\ref{appendix:PSM-code-example}.
\subsubsection{Security Encapsulation and Decapsulation}
Only examining the functional components of the WiFi protocol is not enough to determine whether the entire protocol is secure or not. The reason is that the messages are securely encapsulated in normal transmissions and then sent to the communication channel. The security encapsulation usually includes encryption and authentication. 
Therefore, the security of transmitted messages depends on the security encapsulation mechanism. 

In fragmentation and PSM, security encapsulation is also important.
For our analysis, we follow the \textbf{Section 12.5 Robust Security Network Association (RSNA) protocol} of WiFi specification~\cite{specification}. Among these integrity protocols, because TKIP is deprecated and more secure protocols CCMP and GCMP are recommended~\cite{vanhoef2014advanced}. We thus design the security encapsulation and decapsulation model to represent the CCM Protocol (CCMP) and GCM Protocol (GCMP).   
Our model is based on the \textbf{Section 12.5.3 CCMP} and \textbf{Section 12.5.5 GCMP} of the WiFi protocol specification ~\cite{specification}. 

\begin{figure}[h]
  \centering
  \includegraphics[width=\linewidth]{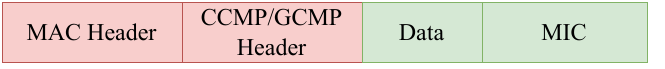}
  \caption{Structure of Security Encapsulated MPDU: the data and MIC fields (green area) are encrypted during transmission; the header fields (red area) are not encrypted.}
  \label{fig:ccmp-gcmp-mpdu}
\end{figure}

The composition of a security-encapsulated MPDU compromises several elements: the MAC header, CCMP or GCMP header, data, and the Message Integrity Code (MIC), as shown in Fig~\ref{fig:ccmp-gcmp-mpdu}. 
MIC is to ensure message integrity. The MAC and CCMP/GCMP headers remain unencrypted. The data and MIC are encrypted to ensure secrecy.  
The MAC header includes various fields, such as the MAC address, sequence number, fragment number, more fragment flag, and so on. 
Conversely, the CCMP/GCMP header incorporates fields including the key ID and packet number. The packet number is used to protect against replay attack~\cite{replay-attacks}. 
When we model the MAC header, we only focus on the related fields and exclude unrelated ones, such as the QoS Control and HT Control fields. The abstraction simplifies the model and reduces the demanding computational requirements. 
Importantly, by retaining all relevant fields, we ensure that the integrity of our security evaluation is not compromised.

\begin{figure}[h]
  \centering
  \includegraphics[width=\linewidth]{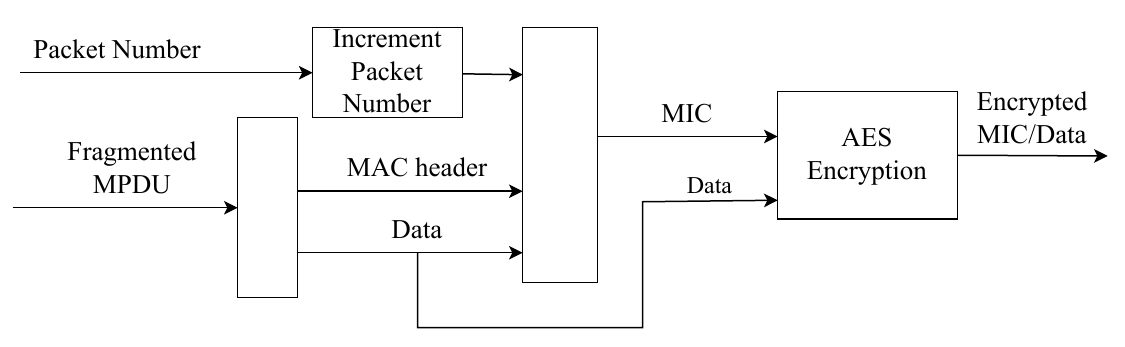}
  \caption{How the encrypted data and encrypted MIC are generated. The MIC is generated from headers and data. }
  \label{fig:ccmp-process}
\end{figure}

We now provide details about the security encapsulation process as shown in Fig~\ref{fig:ccmp-process}. 
For every transmitted MPDU, one must increment the Packet Number (PN), ensuring the generation of a distinct PN. Note that, for a consistent temporal key, the PN must remain unique to preserve security.
The MIC's is generated from the CCMP/GCMP header, the MAC header, and the data frame. The MAC header and PN and data constitute the Additional Authentication Data (AAD). 
Subsequently, the MIC is generated from the AAD.
Once the MIC code is derived, both the data and MIC are encrypted using  AES encryption with the temporal pairwise key.
While the standard specification employs CCM or GCM encryption for the MIC and data, currently, Tamarin 
does not provide support for CCM or GCM encryption. Moreover, our model sets cryptography as perfect. Our model uses Tamarin's built-in symmetric encryption and decryption functions \textsf{senc} and \textsf{sdec} to represent AES encryption. The simplified Tamarin code to implement the encryption is given in Appendix~\ref{appendix:decapsulation-code-example}. And we also provide the packet number check details in Appendix~\ref{appendix:packet-number-check}.
\looseness = -1

\subsection{Properties Extraction}
\label{sec:property-extraction}
Our properties are extracted from the WiFi protocol specification ~\cite{specification} and the WiFi secure requirements. We now
introduce some of the requirements in the specification, as well as the security properties written in the Tamarin code corresponding to the security requirements.

\begin{mybox}{gray}{\textbf{Section 12.5.3.3.7 CCM originator processing}  }
\small
    CCM originator processing provides authentication and integrity of the frame body and the AAD and data confidentiality of the frame body.
\end{mybox}
From the text box before, the message sent under CCMP should keep the confidentiality, authentication, and integrity of data, as well as authentication and integrity of AAD (additional authentication data) generated from the MAC header and CCMP header. 
Therefore, we derive several security properties from such a requirement. The following Tamarin code piece is an example of a rule specifying the frame body confidentiality property. 

\begin{myboxwithouttitle}{blue}
\small
   \textsf{not(} \newline
    \hspace*{2mm} \textsf{Ex \#i \#j msg fragNum seqNum nonce senderID.} \newline
    \hspace*{2mm} \textsf{SenderSendFragment(senderID,msg,fragNum,seqNum,nonce)}  \newline
    \hspace*{2mm} \textsf{@ \#i \& K(msg) @ \#j )}   
\end{myboxwithouttitle}

This property means that if the \textsf{Sender} sends a fragment \textsf{msg} with parameters \textsf{(fragNum, seqNum ..)}, then there is no situation or fact \textsf{K(msg)} allowing the attacker to learn the message \textsf{msg} content.

Besides the essential security properties, including authentication, integrity, and confidentiality, we have integrated more general properties
into our model to ensure comprehensive functionality. 



The following sentence shows one instance of such a requirement for PSM: the STA can only enter the power save mode if the message sent to the AP has the \textit{Power Management} flag set to 1.

\begin{mybox}{gray}{\textbf{Section 11.2 Power management}}
\small
    To change power management mode, a STA shall inform the AP by completing a successful frame exchange that is initiated by the STA.
\end{mybox}

From the above requirement, 
we obtain the following property:

\begin{myboxwithouttitle}{blue}
\small
    \textsf{All \#j apSSID staAddress.} \newline
    \textsf{APKnowDoze(apSSID,staAddress) @ \#j} \newline
    ==> \newline
    \textsf{(Ex \#i. STASendDozeMsg('1',staAddress) @ \#i \& i<j)} 
\end{myboxwithouttitle}

In our Tamarin code translation, if the AP recognizes that a certain STA enters the PSM  at time \textsf{j}, then the event \textsf{STASendDozeMsg} must have occurred before time \textsf{j}.
Meanwhile, the power management flag should be '1'.

After properties extraction and translation to Tamarin code, we get \nProperties{} properties in total. More properties examples are shown in Appendix~\ref{appendix:property-example}. And we also provide a subset of important properties in Appendix~\ref{appendix:properties-table}.

\subsection{Patched-version Model}
\label{sec:patch-version}
Apart from the model construction from specification, we also implement the patches of identified known attacks and our proposed new issues.
Our model identifies two new vulnerabilities and confirms three existing attacks in \cite{vanhoef2021fragment,schepers2023framing}. Two of these attacks are related to the fragmentation process, while one targets the PSM. We implemented the patched model to evaluate its defense against the identified attacks. 

Regarding fragmentation-related attacks, a key vulnerability arises in the CCMP or GCMP encapsulation process. Here, the integrity code is appended after fragmentation. Thus, each fragment's integrity is guaranteed, but the entire frame is not. 
To address this, we introduce an additional integrity check before fragmentation in the patched model. This ensures that upon reassembly of all fragments to one frame at the receiver's end, the entire frame's integrity is verified, protecting it from potential modification during transmission.
The incorporation of whole frame verification in our model is efficient. Since the original fragmentation process already includes the integrity check for each fragment, adding an additional check for the entire frame introduces minimal overhead. 

Another vulnerability that can be exploited for fragmentation-related attacks is that the buffer units are not cleared when the connection terminates, which is also a  vulnerability that can be exploited by attacks against the PSM. 
Moreover, the received fragments or messages are decrypted and stored as buffer units in plain text.
To address this, our patched-version model ensures that the buffer units are cleared once the STA disconnects from the AP.

Another root cause of the attacks is that the \textit{sequence number} of the MAC header is not protected, so it can be altered arbitrarily by an attacker. 
Therefore, adding the protection of \textit{sequence number} is important to defend against these attacks.

\section{Evaluation}
\label{sec:evaluation} 
In this section, we experimentally assess our system to answer the four research questions:
\begin{itemize}
\item \textbf{RQ1. Property Violations:} How many properties are tested in our Tamarin model? How many property violations are detected? 
\item \textbf{RQ2. New Attacks Discovered:} What new attacks are uncovered by our model? 
\item \textbf{RQ3. Impact on Commercial Devices}  What impact do our proposed new issues have on commercial off-the-shelf devices? 
\item \textbf{RQ4. Verification of Existing Attacks:} How many known attacks are verified by our model? Additionally, how does our implementing patched model perform against these known attacks? 
\end{itemize}
\subsection{Properties Violation}
The properties are extracted from WiFi protocol specification~\cite{specification} and verified with Tamarin.
We extract and verify \nProperties{} properties in total and got \nPropertyViolations{} properties violation from them. 
There are situations in which multiple property violations correspond to one vulnerability. 
And some property violations are too minor to represent as a  vulnerability.
For instance, the property ensuring that an AP buffers messages only for the STA in sleep mode, along with the integrity property of the \textit{Power Management} bit, are both falsified. The same underlying cause for both cases is the lack of protection for the \textit{Power Management} field.
Therefore, from the properties violation, we obtain \nVulnerabilities{} WiFi protocol design vulnerabilities. The vulnerabilities are as follows:
\begin{itemize}
    \item The \textit{sequence number} and \textit{retry} field can be altered arbitrarily during transmission.
    \item The \textit{Power Management} field's lack of protection can lead to situations where an Access Point (AP) incorrectly assumes a Station (STA) has entered power save mode despite the STA being in active mode.
    \item The unprotected \textit{More Data} field creates a vulnerability wherein a Station (STA) could mistakenly revert to power save mode before the Access Point (AP) has completed transmitting buffered units.
\end{itemize}


To summarize, the vulnerabilities arise from the absence of protection for specific MAC headers within the security encapsulation of WiFi.  
These fields are not protected or authenticated during transmission; hence, the attacker can manipulate the fields arbitrarily, which will result in several security issues. 

\subsection{New Discovered Attacks}
From the protocol design vulnerabilities, we report \nAttacks{} new attacks. 
\subsubsection{Basic Denial-of-Service Attack}

\begin{figure}[h]
  \centering
  \includegraphics[width=0.8\linewidth]{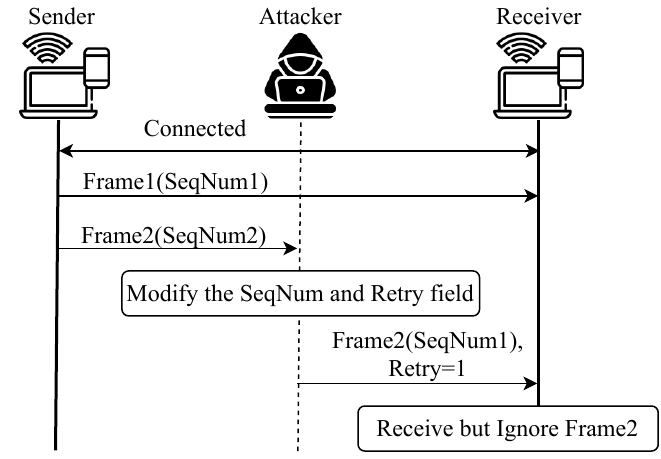}
  \caption{Basic DoS attack. Frame1 with Sequence Number 1 is sent, received, and processed. The Sequence Number of Frame2 is altered to Sequence Number 1, and the Retry field is altered to 1. Then, the receiver mistakes Frame2 as the retransmission of Frame1 and ignores it.}
  \label{fig:base-dos}
\end{figure}

Our formal model verification reveals that the \textit{sequence number} and \textit{retry} fields lack protection, allowing attackers to alter them arbitrarily during transmission. Therefore, we design a DoS attack that makes the receiver ignore the message sent by the sender. In this scenario, the STA and AP are both affected, so we just use the sender and receiver to present them.
The DoS attack is illustrated in Fig.~\ref{fig:base-dos}.

After the connection, the sender starts to send frames to the receiver. \textsf{Frame1(SeqNum1)} is the first frame with \textit{Sequence Number1}, and  \textsf{Frame2(SeqNum2)} is the second frame with \textit{Sequence Number2}. 
In the first step, the first frame is received and processed without issues. 
In the second step, the attacker modifies the second frame's MAC header. The \textit{sequence number} is altered to match the first frame's, and the \textit{retry} field is set to 1. Since the receiver already processed a frame with \textit{Sequence Number1}, it mistakes the attacker-modified second frame for a retransmission of the first frame.This attack can compromise data transmission and open avenues for more sophisticated attacks, potentially leading to severe data loss.


\subsubsection{Denial-of-Service Attack against Fragmentation}
\begin{figure}[h]
  \centering
  \includegraphics[width=0.8\linewidth]{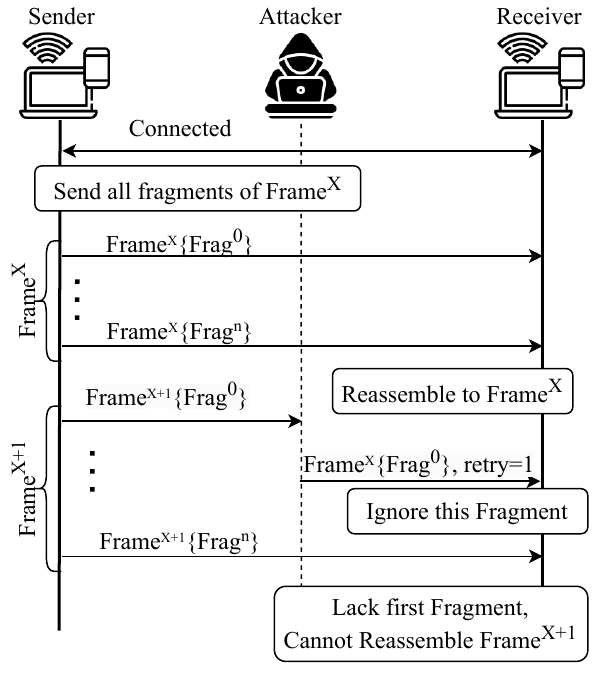}
  \caption{Denial-of-Service Attack against Fragmentation. X denotes the sequence number, and 0 to n denotes the fragment number. The Frame\textsuperscript{X} is successfully fragmented and reassembled on the receiver side. The Frag\textsuperscript{0} of Frame\textsuperscript{X+1} is modified by the attacker, altering the sequence number and retry field. Hence, the receiver ignores the Frag\textsuperscript{0} and cannot reassemble Frame\textsuperscript{X+1} successfully. }
  \label{fig:dos-attack}
\end{figure}

We now introduce an advanced DoS attack targeting the fragmentation process, which is illustrated in Fig~\ref{fig:dos-attack}, building on the basic DoS attack.

In the first step, the sender sends \textsf{Frame\textsuperscript{X}} after the connection. During the process, the frame is fragmented, sent, and reassembled by the receiver successfully. 
In the second step, when the sender needs to send \textsf{Frame\textsuperscript{X+1}}, the attacker modifies the first fragment \textsf{Fragment\textsuperscript{0}} by altering the \textit{sequence number} from $X+1$ to $X$ and altering the \textit{retry} field from 0 to 1, which tricks the receiver into believing that current fragment is duplicate retransmission of a previously received fragment. 
As a result, the reassembly of \textsf{Frame\textsuperscript{X+1}} fails due to the missing \textsf{Fragment\textsuperscript{0}} on the receiver side.

In our tests, we discover that even if only one fragment from a series is missing, the receiver's defragmentation fails. 
By modifying the header field of fragment 0, which is the first fragment and is present in any fragment series, we can affect any fragmented MAC MSDU with this method.

This fragmentation-targeted DoS attack has two benefits for the attacker. First, the attacker only has to modify the header of a single fragment to disrupt the reassembly, effectively causing the entire frame's transmission to fail.
Second, the sender mistakenly believes that the receiver has successfully gotten the first fragment, eliminating any retransmission attempts. Given that a fragmented MSDU only uses immediate acknowledgment, an ACK is sent after each fragment's receipt. 
Furthermore, \textit{sequence number} is not shown in the ACK message. 
If an attacker alters a fragment's \textit{sequence number}, the sender remains believing that the fragment was successfully received. Hence, the sender will not try to retransmit that fragment.

This attack can disrupt the reassembly by only modifying the header of a single fragment. 
This attack not only poses a direct threat to the stability and efficiency of the network but may also cause multi-vector attacks.

\subsection{Impact on Commercial Devices}
To evaluate the effects of our newly discovered attacks on commercial devices, we build a testbed based on the WiFi-framework~\cite{schepers2021framework} and examine \nDevices{} devices, including laptops, phones, network cards, and Internet-of-Things (IoT) devices. Our testing devices are from multiple vendors (such as Qualcomm, MediaTek, Huawei, etc.) and support from WiFi 4 to WiFi 6 generations.

We employ the \textit{Alfa AWUS036ACM} network card on XPS 13 Plus laptop with an Intel i7-1260P CPU, 16 GB DDR3 RAM, and Ubuntu 22.04 system. Devices are assessed using \textit{tcpdump}~\cite{tcpdump2021} and Android Debug Bridge (ADB)~\cite{adb}, serving as Stations. In this setup, the AP acts as the sender, while the STA acts as the receiver. 
\nVulDevices{} commercial devices out of \nDevices{} testing cases are vulnerable to our proposed new issues. The devices and outcomes are detailed in Table~\ref{tab:test-results} of Appendix~\ref{appendix:testing-results}.

From our tests, the devices using the latest WiFi 6 generation exhibit low vulnerability. 
In contrast, the devices using older WiFi generations are all vulnerable. Therefore, we recommend upgrading to a newer WiFi version to enhance device security. More details are discussed in Sec~\ref{sec:discussion}.

\subsection{Existing Attacks Verification}

\begin{table*} 
\centering
\renewcommand{\arraystretch}{1}
\fontsize{7}{7}\selectfont
\caption{The three existing attacks detected by our model, the corresponding patches, as well as the verification results of the original model and the patched model.}
\label{tab:existing-attack}
\begin{tabularx}{\textwidth}{>{\centering\arraybackslash}p{2cm} >{\centering\arraybackslash}p{3.5cm} >{\centering\arraybackslash}p{1.5cm} >{\centering\arraybackslash}p{1cm} >{\centering\arraybackslash}p{3cm} >{\centering\arraybackslash}X} 
\toprule
\textbf{Attacks} & \textbf{Description} & \textbf{Reference} & \textbf{Detected} & \textbf{Patched Version} & \textbf{Remarks} \\
\midrule
Mixed key attack & 
Attack can forge frames by mixing fragments that are encrypted under different keys. & 
\cite{vanhoef2021fragment} & 
\ding{51} & 
Authenticate the entire frame before it is divided into fragments. &
Our model detects frame integrity falsification in the fragmentation model with the key refresh. While the patched-version model verifies the same integrity property.\\
\addlinespace
Poisoning the fragment cache attack & 
An adversary injects fragments into memory with MAC spoofing. & 
\cite{vanhoef2021fragment} & 
\ding{51}  & 
Clear the cache when one device disconnects from the current AP. &
Our model identifies frame integrity falsification in the fragmentation model under MAC spoofing. While the patched-version model verifies the same integrity model.\\
\addlinespace
Leaking frames from WiFi queue & 
The attacker forces an AP to queue frames, changes the security association, and leaks the queued frames. & 
\cite{schepers2023framing} & 
\ding{51} & 
Clear the cache when one device disconnects from the current AP. &
In the PSM model with MAC spoofing, we found that the secrecy of the queued buffer is falsified. The same property in the patched version is verified. \\
\bottomrule
\end{tabularx}
\end{table*}

Our model detects \nKnownAttacks{} known attacks from prior research \cite{vanhoef2021fragment,schepers2023framing}. In addition, we design, implement and validate the patched-version model of these existing attacks. In the patched-version model, we enhance security by integrating a whole frame integrity check and implementing cache-clearing mechanisms. 
The results show the effectiveness of the patched-version model. The details are shown in Table~\ref{tab:existing-attack}.

\section{Discussion}
\label{sec:discussion}
\textbf{Why Choosing Fragmentation and PSM. } 
At first, we systematically and exhaustively review the previous works on formal verification of WiFi protocol. The works include the formal verification of key-exchanged handshakes, such as ~\cite{cremers2020formal}, and the formal verification of Voice over WiFi protocol ~\cite{lee2022vwanalyzer}. However, there is no work verifying the security of fragmentation and PSM in WiFi protocol. Therefore, our work aims at filling this void.

Second, attacks on these two parts have been discovered in recent years, which have caused severe impacts such as user information leakage. Vanhoef et al.~\cite{vanhoef2021fragment} identified weaknesses in the fragmentation part of WiFi protocol. 
The weakness arises from the whole frame is not authenticated, and only a series of fragments are authenticated separately. 
Therefore, the researchers came up with mixed-key attacks which can leak user information. 
Schepers et al.~\cite{schepers2023framing} discovered vulnerabilities in PSM of WiFi protocol, as the buffered units are not cleared after disconnection.
Also, they came up with an attack that can leak the user buffer content, which may lead to information leakage. 
Because the attacks are severe, we want to build a Tamarin model that can detect these types of vulnerabilities.

Ultimately, the fragmentation and PSM components both use the frame buffer mechanism in WiFi protocol. 
During the fragmentation process, an MSDU is divided into multiple MPDUs to be sent; on the receiver side, the MPDUs sent earlier will be decrypted and buffered.
In the PSM mechanism, the AP  will buffer the messages that need to be sent to the sleeping non-AP STA.
In addition, the MAC header fields related to fragmentation and PSM are both not protected and authenticated, and starting from this point, it will lead to several weaknesses of these two components. 
In conclusion, fragmentation and PSM have two similarities: buffer mechanism and unprotected MAC header fields. The two components are also relatively important WiFi functions related to the buffer mechanism and unprotected MAC header. Thus we implement the formal verification of these two parts in our work.

\noindent \textbf{Analysis of Devices Testing Results.} 
We evaluate \nDevices{} devices spanning WiFi generation 4 to 6, supported by multiple vendors. As illustrated in Table~\ref{tab:test-results}, \nVulDevices{} of these devices were found vulnerable, proving the wide range and significance of our identified issues.
Upon closer checking of the non-vulnerable devices, we observed a common fact: they all utilize WiFi 6 and operate on Android 11 or 12.
This suggests that the WiFi generation 6 devices are more robust with respect to the issues we have identified. However, the presence of vulnerable devices with both WiFi 6 and Android 12 shows that the devices with the new versions can still be attacked.
The devices with Android 10 and WiFi 5 generation, as well as the earlier generation devices, are all vulnerable. Therefore, it is necessary to upgrade the WiFi version and the operating version. 
Furthermore, our tested laptops, network card, and Internet-of-Things (IoT) devices are also vulnerable, highlighting the issues' wide-range impact.

\noindent \textbf{Ethical Consideration \& Responsible Disclosure.}  We have conducted commercial device testing in a controlled environment, only affecting our own controlled devices. We only performed the experiments by sending messages with their constraints. Our purpose is to evaluate the effectiveness of our proposed new attacks, not to cause any damage. 
For the new issues, we have responsibly disclosed the findings to all the vendors and are cooperating with them for additional needed information. As the discussed issues can, at max, cause denial-of-service, the vendors consider them as having a low-security impact. Nonetheless, some of the vendors are coordinating further on defenses. We thank the vendors for looking into the issues and taking the user's security and privacy very seriously.




\section{Related Work}
Work-related to ours can be classified in 
two broad categories: (i) Attacks on WiFi and (ii) Formal verification of protocols.
\subsection{Attacks on WiFi}

Over the years, the WiFi security protocol has evolved from WEP to WPA, then WPA2, and now WPA3. Very well-known attacks are the ones 
against 
the WiFi key exchange handshake.
KRACK~\cite{vanhoef2017key,vanhoef2018release} is one such attack. It exploits nonce reuse to infer the encryption key. The attack causes a repeat of the third message of the WPA2 four-way handshake process.
Even though WPA3 was then introduced, it has not addressed all the vulnerabilities of WPA2~\cite{kohlios2018comprehensive}. 
Vanhoef et al. ~\cite{vanhoef2020dragonblood} showed that attackers could make WPA3 Access Points use weaker encryption by changing the SAE handshake process or even making devices downgrade to WPA2. Also, Chatzoglou et al. ~\cite{chatzoglou2022your} showed attacks that disrupt WiFi service via DoS on WPA3-SAE, which could affect many APs.

Beyond attacks on key exchange handshakes of WPA2 or WPA3, there are also vulnerabilities in other aspects of the WiFi protocol. 
The Kr00k~\cite{kr00k} attack was found in 2019. The researchers found that some encrypted traffic could be decrypted without authorization. This happens when an STA disconnects, forcing the WiFi hardware chip's key to reset to zero, which could leak data.
Vanhoef et al. ~\cite{vanhoef2021fragment} identified weaknesses in the aggregation and fragmentation components of the WiFi protocol. These vulnerabilities arise because the A-MSDU (Aggregation MSDU) header field and sequence number header field are not adequately protected. By exploiting these weaknesses, researchers have proposed mixed-key attacks and cache poisoning attacks under MitM and BEAST threat models.
Schepers et al. ~\cite{schepers2023framing} discovered vulnerabilities and attacks against the PSM and framing queue components of the WiFi protocol, which can expose buffers under the MAC spoofing threat model.
Such vulnerabilities stem from the fact that the queued frames are stored in plaintext. 
Schepers et al. have also shown that DoS attacks are possible by deceiving an AP into believing a client is in sleep mode.

We have observed that components of the WiFi protocol outside of the RSNA can also lead to serious issues, including the leakage of sensitive information. However, past efforts to verify the WiFi protocol have primarily focused on the authentication process and key-exchange handshake. They have largely overlooked other general functional components like fragmentation and PSM. Our research aims to address this gap by focusing on the fragmentation and PSM components of the WiFi protocol. Furthermore, though the previous works on finding attacks in functional components provide notable results, they do not use any systematic approach: 
(i) the attacks are found via manual analysis, which has limitations because of the protocol complexity; (ii) there is no formal verification of protocol issues. 


\subsection{Formal Verification}

Several approaches have been proposed to use formal methods to analyze the security of network protocols.
Hussain et al.~\cite{hussain20195greasoner} proposed 5GReasoner, a systematic methodology combining two model checkers and a cryptographic verifier, for the formal analysis of the security of the control-plane protocols spanning across multiple layers of the 5G protocol stack. 
Basin et al. ~\cite{basin2018formal} focused on 5G Authentication and Key Agreement protocol, conducting a systematic security verification and suggesting security patches to fix vulnerabilities found from the analysis.
Wang et al.~\cite{wang2021mpinspector} proposed MPInspector, a tool designed to verify the implementation of Messaging protocols.
Shi et al. ~\cite{shiformal} used Tamarin to verify the security of the BLE Secure Connection protocol. They relaxed the perfect cryptography in Tamarin, enabling their framework to disclose low-entropy key leakage attacks. 
Jacomme et al.~\cite{jacomme2023comprehensive} utilized SAPIC+~\cite{cheval2022sapic+} to verify the EDHOC protocol, which is a lightweight key exchange protocol.

In the area of WiFi verification, Cremers et al. ~\cite{cremers2020formal} were the first to verify the WiFi protocol, focusing on the WPA2 four-way handshake, the group-key handshake, and the WNM sleep mode. They verified the KRACK attack~\cite{vanhoef2017key} by relaxing the perfect cryptography assumption in Tamarin. However, their work did not cover the fragmentation and power save mode of WiFi, limiting their verification to the KRACK attack; thus, their approach is not able to discover other attacks such as FragAttacks~\cite{vanhoef2021fragment}.

Our work differs from previous work in the area of formal analysis of WiFi protocol in that it is the first to focus on the verification of more functional elements of the WiFi protocol, such as the fragmentation and power-save mode. 
Moreover, we consider modeling the MAC spoofing threat model with Tamarin. We incorporate a model of MAC spoofing threats to expand the range of potential attacks our system can verify.
In addition, we propose the segment-based methodology to construct the Tamarin model.
In the previous work~\cite{wu2022formal} on Bluetooth formal verification, the authors divide the protocol flow into different linear procedures and model them separately as modules, whereas in our case, the functional protocol interacts with the WiFi security protocol, creating complex protocol interactions. For example, one frame may be in the process of authentication, fragmentation, encryption, and integrity code addition. To resolve this, we propose the segment-based design where each protocol is a segment interacting with the other segments. On a high level, our segment modeling captures the complex multi-protocol interactions compared to the previous work, where the protocol flow is divided linearly one after the other.


\section{Conclusion \texorpdfstring{\&}{} Future Work}
In our work, we present the formal verification of WiFi functional components, fragmentation, and Power Save Mode (PSM). We design a \emph{segment-based} method to build the complex Tamarin model to achieve this. Additionally, we extend the scope of verification by introducing a practical threat model involving MAC spoofing.
As for the results, we verify \nProperties{} properties in total, which are extracted from WiFi protocol specification. 
We then identify and present \nAttacks{} new issues of WiFi protocol from the verification findings.
Furthermore, we test our proposed new issues on \nDevices{} commercial devices, and \nVulDevices{} devices are vulnerable, showing the wide-range impact of the issues.

\noindent \textbf{Future work.} For future work, we will verify other functional components of the WiFi protocol, such as the MSDU Aggregation mechanism and the Block ACK mechanism. 
We would also develop new verification methods that can verify the complete WiFi protocol.
\printbibliography
\appendix
\section{Model Construction Tamarin Code Example}
\label{appendix:code-example}

\subsection{Security Encapsulation and Decapsulation Code Example}
\label{appendix:decapsulation-code-example}
In the following section, we present a simplified piece of the Tamarin code that models the security encapsulation process. 
In our comprehensive modeling, several details related to the adjustments and configurations of header fields are also diligently accounted for.

\begin{myboxwithouttitle}{blue}
\small
\textsf{rule SendFragment}: \newline
  \hspace*{2mm} \textsf{let} \newline
   \hspace*{4mm} \textsf{encryptedMsg = senc(message,key)} \newline
   \hspace*{4mm}  \textsf{MICcode = MIC(<MACheader,SecurityHeader,message>)} \newline
   \hspace*{4mm} \textsf{enMICcode = senc(MICcode,key)} \newline
  \hspace*{2mm} \textsf{in} \newline
   \hspace*{4mm} \textsf{[ Fr(~message),} \newline
   \hspace*{4mm}  \textsf{SenderState(senderID,MACheader,SecurityHeader,key)]} \newline
 \hspace*{4mm} -- \textsf{[ SendFragment(senderID,message,} \newline
   \hspace*{4mm} \textsf{MACheader,SecurityHeader)} ]-> \textsf{ [ Out(<MACheader,}\newline
   \hspace*{4mm} \textsf{SecurityHeader, encryptedMsg, enMICcode>) ]}
\end{myboxwithouttitle}

The \textsf{encryptedMsg} and \textsf{enMICcode} are encrypted with a key and Tamarin's built-in symmetric encryption function \textsf{senc}. The \textsf{SecurityHeader} means the CCMP or GCMP header is in MPDU format.
We define a \textsf{MIC} function to generate MIC code, which will be used to verify the message integrity. Sending a message also follows the format in Figure~\ref{fig:ccmp-gcmp-mpdu} as \textsf{Out(<MACheader,SecurityHeader, encryptedMsg, enMICcode>)}.
The security decapsulation on the receiver side is the converse process of the security encapsulation, we put the Tamarin code of decapsulation in Appendix~\ref{appendix:decapsulation-code-example}.

\label{appendix:decapsulation-code-example}
The decapsulation process functions as the inverse process of the security encapsulation procedure. We provide a simplified code example to illustrate the decapsulation mechanism here. 
Upon receipt of a message structured as \textsf{In(<MACheader,CCMPheader, encryptedMsg, enMICcode>}, the recipient uses its inherent key to decrypt both the encrypted message and the encrypted MIC. To accomplish this decryption, we employ Tamarin's integrated symmetric decryption function, \textsf{sdec()}.
Subsequent to decryption, the receiver focuses on MIC calculation using the received message, represented as \textsf{calMIC = MIC(<MACheader,CCMPheader,message>)}. The integrity of the header and data is preserved if, and only if, the computed MIC aligns with the decrypted MIC obtained from the received message. 

\begin{myboxwithouttitle}{blue}
\small
\textsf{rule RecFragment:} \newline
\hspace*{2mm}  \textsf{let} \newline
\hspace*{4mm}     \textsf{message = sdec(encryptedMsg,key)} \newline
\hspace*{4mm}     \textsf{MICcode = sdec(enMICcode,key)} \newline
\hspace*{4mm}     \textsf{calMIC = MIC(<MACheader,CCMPheader,message>)} \newline
\hspace*{2mm}   \textsf{in} \newline
\hspace*{4mm}     \textsf{[ In(<MACheader,CCMPheader, encryptedMsg, } \newline
\hspace*{4mm}     \textsf{enMICcode>), ReceiverState(receiverID,key)]} \newline
\hspace*{4mm}   --\textsf{[ Eq(MICcode,calMIC),} \newline
\hspace*{4mm}      \textsf{RecFragment(senderID,message,} \newline
\hspace*{4mm}     \textsf{MACheader,CCMPheader) ]}-> \newline
\hspace*{4mm}     [ ... ]
\end{myboxwithouttitle}

\subsection{PSM Code Example}
\label{appendix:PSM-code-example}

Here, we provide a simplified example of how we translate the step: "AP receives the message with Power Management flag equal to 1, AP sends the ACK message and stores the doze state of STA".

\begin{myboxwithouttitle}{blue}
\small
  \textsf{rule APRecDozeMsg:} \newline
   \hspace*{2mm} \textsf{let} \newline
     \hspace*{4mm} \textsf{msg = sdec(encryptMsg, key)} \newline
     \hspace*{4mm} \textsf{MIC = sdec(encryptMIC, key)} \newline
     \hspace*{4mm} \textsf{mic = MIC(<staAddress, msg>)} \newline
     \hspace*{4mm} \textsf{ack = 'ack'} \newline
     \hspace*{4mm} \textsf{enAck = senc('ack', key)} \newline
    \hspace*{2mm} \textsf{in} \newline
    \hspace*{4mm} \textsf{[ In(<<pwrMgmt,staAddress>, encryptMsg, encryptMIC>),} \newline
    \hspace*{4mm} \textsf{APState(apSSID,key)} ] \newline
     \hspace*{4mm} --- \textsf{[ Eq(mic,MIC), Eq(pwrMgmt,'1') ]}-> \newline
     \hspace*{4mm} \textsf{[ APStateKnowDoze(apSSID,key,)} \newline
     \hspace*{4mm} \textsf{APSaveDoze(apSSID,staAddress), Out(enAck) ]} 
\end{myboxwithouttitle}
The right-hand fact \textsf{In(<<pwrMgmt,staAddress>, encryptMsg, encryptMIC>)} means that the AP receives the message from STA. And action \textsf{Eq(pwrMgmt,'1')} is to verify if the power management field equals to 1. The right-hand fact \textsf{APSaveDoze(apSSID} \textsf{,staAddress)} means that the AP knows the doze state of STA. And \textsf{Out(enAck) } means that the AP sends the ACK message to STA. 
In conclusion, this rule shows that when the AP receives the message which \textit{Power Management} field equals to 1 from some STA, it will store the doze state of the STA and send the ACK message back.

\subsection{Packet Number Check}
\label{appendix:packet-number-check}

To protect against replay attacks, packet numbers are utilized in WiFi CCMP and GCMP.
Each time the sender transmits a message, the PN is incremented by one. This ensures that every message encrypted with the same pairwise key has a unique PN. On the receiving end, the system checks if the PN of a new message is greater than the previous one. If this is not the case, the message is discarded.
The code below shows the rule for the packet number-checking mechanism using Tamarin at the receiver's end.

\begin{myboxwithouttitle}{blue}
\small
\hspace*{2mm}  [ In(<fragNum,seqNum,PN,encryptedMsg,enMICcode>)\newline
\hspace*{2mm}  , ReceiverSecurityState(receiverID,rPN,key)  ] \newline
\hspace*{2mm}  --[Eq(PN, rPN+1) ]-> \newline
\hspace*{2mm}  [ Process(fragNum,seqNum,PN,encryptedMsg,enMICcode) ]
\end{myboxwithouttitle}
In this rule before, \textsf{PN} means packet number. At the receiver side, the receiver stores the PN of the last previously received message, denoted as \textsf{rPN}. The receiver will verify that the newly received PN is equal to the PN of the previous message plus 1 (i.e., \textsf{Eq(PN, rPN+1)}). If this condition is true, the message will be further processed; if not, the message will be discarded.

\section{Security Properties Examples}
\label{appendix:property-example}
\begin{mybox}{gray}{\textbf{Section 10.4 MSDU and MMPDU fragmentation} }
\small
     Once a fragment is transmitted for the first time, its frame body content and length shall be fixed until it is successfully delivered to the immediate receiving STA.
\end{mybox}

This above message reveals that the sent fragment should be provided integrity, which means that the message sent by the sender and the message received by the receiver should remain the same. We encode this requirement to a Tamarin property below:
\begin{myboxwithouttitle}{blue}
\small
    \textsf{All \#j msg fragNum seqNum nonce receiverID. \newline
    ReceiverRecFrag(receiverID,msg,fragNum,seqNum,nonce) \@j \newline
    ==>  \newline
    (Ex \#i senderID fragNum2 seqNum2 nonce2.  \newline
    (SenderSendFragment(senderID,msg,fragNum2, \newline
    seqNum2,nonce2) @i) \& i<j)} 
\end{myboxwithouttitle}


\section{Important Properties}
\label{appendix:properties-table}
We present a subset of important properties in Table~\ref{tab:properties}.
\begin{table}[h]
\centering
\caption{Subset of the Evaluated Properties}
\label{tab:properties}
\begin{tabular}{|l|}
    \hline
    \textbf{Properties} \\ \hline
    Integrity of Fragment for Fragmentation        \\ \hline
    Integrity of Frame for Fragmentation         \\ \hline
    Integrity of Fragment Number for Fragmentation           \\ \hline
    Integrity of Sequence Number for Fragmentation  \\ \hline
    Integrity of Nonce for Fragmentation  \\ \hline
    Integrity of Power Management Field for PSM \\ \hline
    Integrity of STA Address for PSM \\ \hline
    Secrecy of Fragment for Fragmentation          \\ \hline
    Secrecy of Frame for Fragmentation  \\ \hline
    Secrecy of Nonce for Fragmentation  \\ \hline
    Secrecy of Key for Fragmentation \\ \hline
    STA enter sleep state after receiving ACK \\ \hline
    AP stores messages after STA sleeping \\ \hline
    The authentication should happen after association for PSM \\ \hline
    The AP should send buffers after it stores buffers \\ \hline
    Secrecy of Message for PSM \\ \hline
    Secrecy of Key for PSM \\ \hline
\end{tabular}
\end{table}

\FloatBarrier
\section{Testing Devices and Results}
\label{appendix:testing-results}
 Details of the tested devices and whether they are vulnerable to our attack are detailed in Table~\ref{tab:test-results}.
 
\FloatBarrier
\begin{table*}[h]
    \centering
    \caption{Commercial Devices Test Results.}
    \label{tab:test-results}
    \begin{tabular}{lcccc}
        \toprule
        Device & Chipset Vendor & WiFi Generation & OS Version & If Vulnerable \\
        \midrule
        Dell XPS 13 plus & Intel & WiFi 5 & Ubuntu 22.04 & \ding{51} \\
        Mi Laptop 15.6 & RealTek & WiFi 5 & Ubuntu 22.04 & \ding{51} \\
        TL-WN722N v2 & TP-Link & WiFi 4 & Ubuntu 22.04 & \ding{51} \\
        Huawei P8 Lite & Huawei & WiFi 4 & Android 5.0 & \ding{51} \\
        HTC One E9PLUS  & MediaTek & WiFi 4 & Android 5.0 & \ding{51} \\
        LG G3 & Qualcomm & WiFi 4/5 & Android 6.0 & \ding{51} \\
        Nexus 6 & Qualcomm & WiFi 4/5 & Android 7.1 & \ding{51} \\
        Honor 8X & Huawei & WiFi 5 & Android 8.1 & \ding{51} \\
        Huawei Y5 Prime & Huawei & WiFi 4 & Android 8.1 & \ding{51} \\
        LG Velvet 5G & Qualcomm & WiFi 5 & Android 10  & \ding{51} \\
        OnePlus 7T & Qualcomm & WiFi 5 & Android 10  & \ding{51} \\
        Xiaomi 12T & MediaTek & WiFi 6 & Android 12 & \ding{51} \\
        OnePlus 8T+ & Qualcomm & WiFi 6 & Android 11 & \ding{55} \\
        OnePlus 9 Pro & Qualcomm & WiFi 6 & Android 12 & \ding{55} \\
        Motorola Edge 39 Pro & Qualcomn & WiFi 6 & Android 12 & \ding{55} \\
        IPhone X & Apple & WiFi 5 & IOS 15.5 & \ding{51} \\
        Amazon Smart Plug & Unknown & WiFi4 & Unknown & \ding{51} \\
        \bottomrule
    \end{tabular}
    \label{tab:your_label}
\end{table*}

\end{document}